\documentclass[draft]{agujournal2019}
\usepackage{url} 
\usepackage{lineno}
\usepackage[inline]{trackchanges} 
\usepackage{soul}
\usepackage{gensymb} 
\usepackage{makecell}
\draftfalse

\journalname{JGR: Machine Learning and Computation}

\begin{document}

%
%


\title{Acceleration of horizontal numerical advection for atmospheric modeling through surrogate modeling with temporal coarse-graining}

%
%




\authors{Manho Park\affil{1}, Christopher V. Rackauckas\affil{2}, Christopher W. Tessum\affil{1}}


\affiliation{1}{The Grainger College of Engineering, Department of Civil and Environmental Engineering, University of Illinois Urbana-Champaign}
\affiliation{2}{Computer Science and Artificial Intelligence Laboratory, Massachusetts Institute of Technology, Cambridge, MA, USA}




\correspondingauthor{Christopher W. Tessum}{ctessum@illinois.edu}



\begin{keypoints}
\item We developed a machine-learned advection solver with the same spatial resolution as the baseline simulation (rather than lower resolution). 
\item Our machine-learned solver shows good generalization performance even when trained with a limited dataset. 
\item Our solvers offer favorable tradeoffs between speed and accuracy, gaining 10$\times$ speedup for every $r^2=0.26$ loss in accuracy. 
\end{keypoints}

%
%

%
%


\begin{abstract}
Machine-learned surrogate modeling of advection may accelerate geoscientific models, but existing approaches have either achieved limited speedup or have sacrificed spatial resolution compared to the model they are trained to emulate. 
We developed a machine-learned solver that speeds up advection simulations without sacrificing spatial resolution through the use of temporal coarse-graining, where the model is trained to take larger integration steps than dictated by the Courant-Friedrich-Lewy (CFL) condition. 
Our solver framework includes a convolutional neural network that takes concentrations and CFL numbers as inputs and outputs mass flux. 
Our solvers emulate 10-day ground-level horizontal advection simulations with r$^2$ values against the baseline ranging from 0.60--0.98 with temporal coarsening factors of 4 to 32 times the baseline integration time step. 
Speed increases and accuracy decreases with increased coarsening, with $r^2 = 0.24$ in accuracy lost for every factor of 10 gained in speed, reaching a maximum 92$\times$ speedup while maintaining $r^2 = 0.60$.
We deliberately trained our solvers only on January ground-level wind data to examine their ability to generalize across seasons and vertical heights. 
The 4$\times$-coarsened learned solver successfully reproduces simulations over 72 vertical levels. 
The 8$\times$--16$\times$ solvers (but not 32$\times$) emulate most vertical levels. 
The learned solvers also generalize well across seasons, except for instabilities in June and October.
With additional fine-tuning, these learned solvers could be appropriate for operational use where trading accuracy for speed could be advantageous, such as in screening tools, in ensemble simulations, or with data assimilation.

\end{abstract}

\section*{Plain Language Summary}
The computer simulation of moving particles and chemicals in air is important for the study of study air quality, weather, and climate. 
Existing simulation programs take a long time to run, causing delays in timely forecasting and research. 
We developed a machine learning-based transport simulator that is up to 100 times faster than the standard algorithm used in a popular air quality model. 
We trained our simulator using ground-level wind data that is typically slower than the wind at high altitude. 
We tested our simulator using wind data from different seasons and different altitudes to evaluate whether the simulator can work with unseen conditions. 
The machine learning simulators show promising capabilities in working on unseen conditions, although they usually work better with a wind velocity range that is close to the training dataset. 
This study strengthens the foundation for the acceleration of transport simulation through machine learning.

%
%


\section{Introduction}\label{sec:intro}
The advection equation represents the movement and conservation of physical quantities such as mass, energy and momentum, three canonical quantities in fluid dynamics \cite{street1996elementary}. 
Its numerical solution is an essential component of the simulation of fluid dynamics-related behaviors in geoscientific models, and is critical for understanding, for example, the spatio-temporal distribution of moisture for weather forecasting or atmospheric composition for air quality modeling.

One major challenge in simulating numerical advection in geoscientific models is its computational cost. 
Advection is often the most expensive operation in geoscientific models such as the Earth Atmosphere Model of Exascale Energy Earth System Model (E3SM EAM) or Community Multiscale Air Quality Model (CMAQ) \cite{zhang2018impact, efstathiou2024enabling}), or the second-most expensive module in the chemical transport models such as GEOS-Chem and the Comprehensive Air Quality Model with Extension (CAMx) \cite{eastham2018geos, cao2023gpu}. 
This limits the number and scale of simulations that can be run, especially when high resolution is required to produce spatially-detailed predictions or accurate simulation outputs. 
At the same time, high computational cost is a barrier to improving the accuracy of the simulation. 
The truncation error, the difference between the numerical and the exact solution, decreases as the spacing of the grid decreases \cite{durran2010numerical}. 
Fine grid spacing can also allow explicit modeling of fine-scale processes such as clouds \cite{donahue2024exascale}. 
However, finer resolution requires a larger number of grid cells to solve the mass balance equation and a shorter time step to meet the Courant-Fredrichs-Lewy (CFL) criteria \cite{courant1928partiellen} for advection, causing a superlinear growth in computational cost. 
Therefore, spatial resolutions typically used in geoscientific models are the result of a compromise between the desire for detail and accuracy and the availability of time and computational resources. 

Recent work has explored the use of machine-learning as one way to reduce the dimensionality of the computational domain for numerical advection by coarse-graining (i.e. reducing the spatiotemporal resolution of the model domain), thereby reducing the number of computational operations required. 
\citeA{kochkov2021machine} demonstrated 86 times speedup in a 2-D computational fluid dynamics simulation through the use of spatial coarse-graining, which built on their previous work on 1-D discretization \cite{bar2019learning}. 
Likewise, \citeA{zhuang2021learned} demonstrated a convolutional neural network-based advection solver that reduced the spatial dimension by the factor of 4 while maintaining the comparable accuracy. 
\citeA{stachenfeld2021learned} performed both spatial and temporal coarse-graining of a 3-D turbulence model and achieved 1000$\times$ speedup, although this speedup was gained from the use of hardware acceleration as in addition to a decrease in the number of computational operations required. 
Our previous work \cite{park2024learned} built on the work of \citeA{zhuang2021learned} and performed spatiotemporal coarse-graining, confirming that the learned solver approach can be applied in a geoscientific modeling context with realistic wind fields. 
\citeA{benson2025atmospheric} demonstrated highly accurate emulation of CO$_2$ transport (r$^2$ = 0.99 over 90 days), however only minimal computational speedup (1.6$\times$) was obtained, even though they combined spatio-temporal coarse-graining with hardware acceleration. 

Traditional spatial discretization schemes are designed to represent local spatial gradients, and this locality is what gives rise to the CFL constraint. 
In contrast, It is possible for machine-learned solvers to take explicit time steps that are larger than the CFL condition because machine learned-solvers can be trained to minimize global simulation error, and as such they are not subject to any fundamental constraints on spatial resolution or time step size. 
Flexibility from the CFL restriction can offer reduction of computational expense. 

However, the use of spatial coarse-graining with machine learning for advection has two major caveats. 
First, coarse-graining while maintaining accuracy is not something that can only be done with machine learning. 
\citeA{mcgreivy2024weak} found that the reported speedup factors in many machine-learning fluid dynamics studies could also be gained using a more advanced traditional numerical method instead of a machine learning solver. 
For example, if a high resolution finite volume method can be replaced using either a low-resolution convolutional neural network \cite{kochkov2021machine} or a low-resolution pseudo-spectral method and get the same speedup, then there is little reason to use the learned solver. 
Second, spatial coarse-graining is not ideal in the context of geoscientific models because the models already have a coarser representation of space than what would be ideal for many use cases, and as such, additional spatial coarsening would degrade the relevance of their outputs. 
For example, the use of coarse spatial resolution in air quality modeling can result in the underestimation of the total amount of pollution exposure \cite{li2016influence} and of the disparity in exposure by race-ethnicity \cite{paolella2018effect}. 

For these reasons, there is a clear use for a machine learning solver that can achieve computational acceleration without compromising spatial resolution. 
The question is whether it is possible to meaningfully speed up advection simulation without spatial coarse-graining. 
There is limited prior literature on this topic: \citeA{chen2024decomposing} demonstrated the machine learning emulation of advection and convection from a numerical weather forecast model in the baseline spatial resolution, but did not report computational speedup. 
In our previous work \cite{park2024learned}, the learned solvers with unmodified spatio-temporal resolution were slower than the reference method used, and they failed to produce stable simulations when temporal coarse-graining was applied without spatial coarse-graining. 

Here, we leverage the fact that, in the context of geoscientific modeling, the temporal resolution of the results is often much less important than the spatial resolution: models typically take computational time steps of seconds to minutes, but archive outputs at a frequency of an hour or longer, providing an opportunity to decrease temporal resolution without sacrificing the relevance of the results, if it is possible to do so with an acceptable level of accuracy. 
We also take advantage of a specialized software package for small neural networks \cite{simplechains} which uses stack-allocated operations to be 5--10 times faster than standard deep learning packages for the small hidden layer operations of our design. 
Below, we demonstrate a machine-learned advection operator that achieves acceleration through temporal coarse-graining alone (without any spatial coarse-graining) and evaluate its performance. 

\begin{figure}
\includegraphics[width=1.0\textwidth]{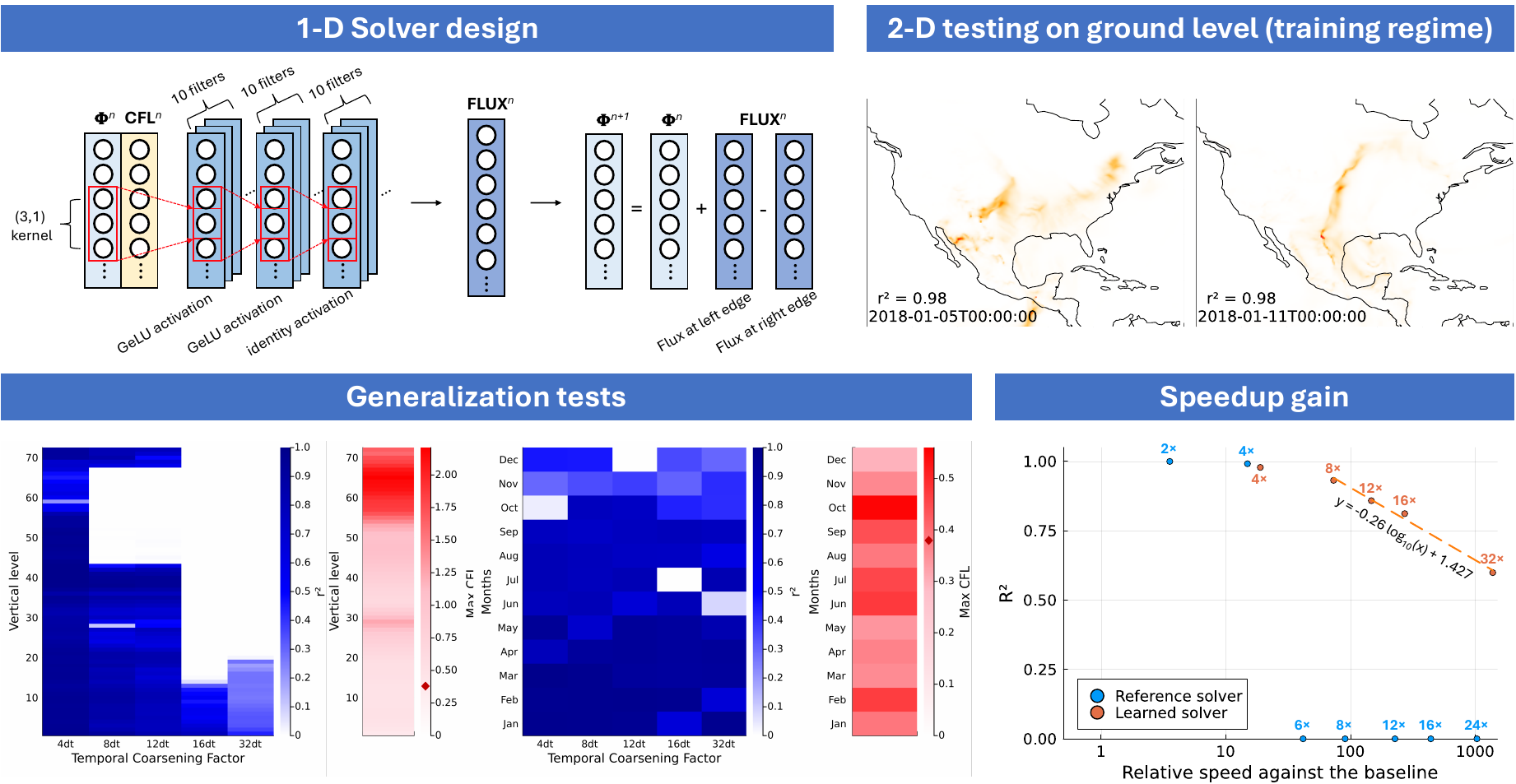}
\caption{Overview of the methods and key results of this study.}
\label{fig:summary}
\end{figure}

The organization of this paper is as follows (represented graphically in Figure \ref{fig:summary}): Section~\ref{sec:methods} explains the data preparation, solver design, training and evaluation protocols. Section~\ref{sec:results} shows the performance, generalization capability, and computational speedup of our solver. 
Section~\ref{sec:discussion} discusses major strengths and limitations of this work and suggests the potential application areas for this approach and future research directions.

\section{Materials and Methods}\label{sec:methods}
We designed and trained a convolutional-neural-network-based advection solver using 1-D advection data generated with ground-level velocity fields over continental North America (130$^\circ$W--60$^\circ$W, 9.75$^\circ$N--60.0$^\circ$N). 
We trained solvers at multiple levels of temporal coarsening (as compared to the timestep used by the reference solver to generate the training data) to explore tradeoffs between accuracy and computational speed. 
We evaluated the solver performance using the 2-D advection dataset in the same spatio-temporal domain and determined the optimal neural network parameters. 
Then, we tested the generalization capacity of the machine-learned solvers to different seasons and vertical levels. 

\subsection{Solver design}
\begin{figure}
\includegraphics[width=1.0\textwidth]{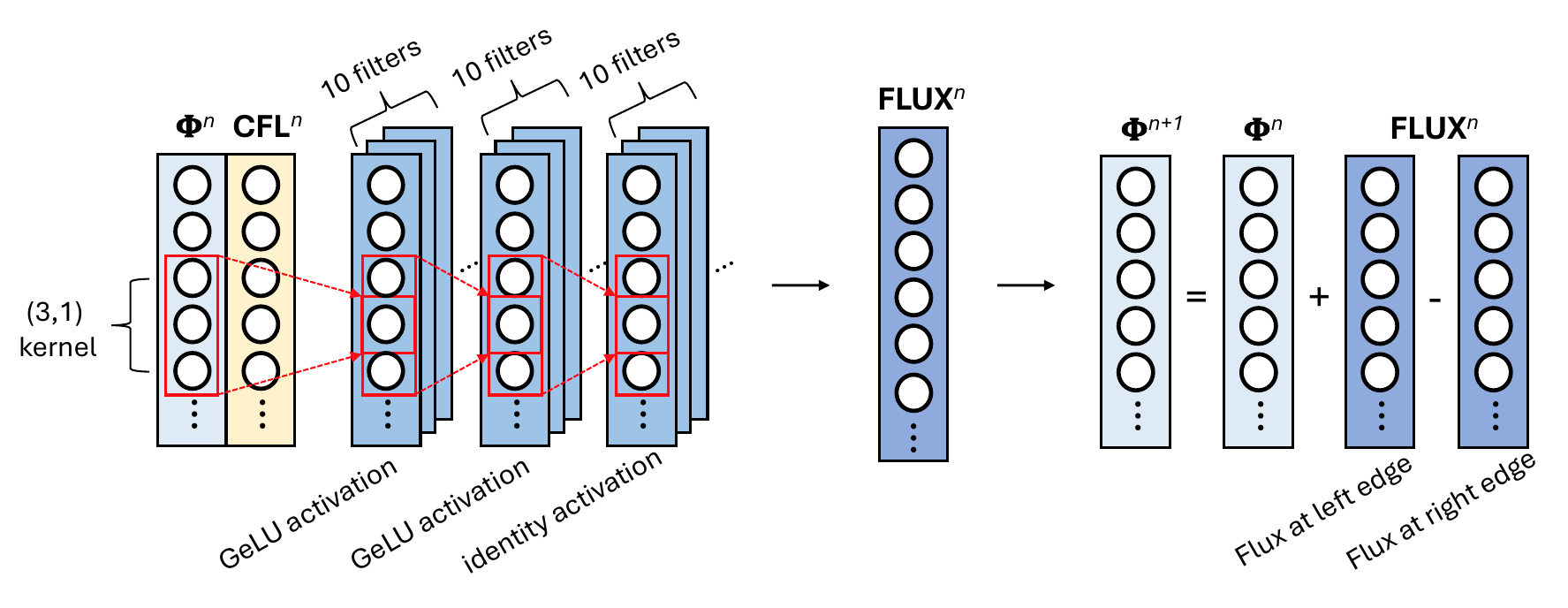}
\caption{A graphical representation of the flux-form solver developed in this study. $\Phi$ is scalar value (i.e. concentration), CFL is CFL number, and FLUX is mass flux at cell edges. Superscript $n$ denotes $n$th time step.}
\label{fig:solver}
\end{figure}

Figure~\ref{fig:solver} visualizes the solver formulation and the process we used to compute concentrations at sequential time steps. 
We built a convolutional neural network-based solver that accepts the current concentration and CFL number for a grid cell of interest and the same information for its neighbors, and outputs the mass flux across the cell boundary on its left side. 
The CFL number was calculated using Equation~\ref{eq:cfl}:

\begin{equation}\label{eq:cfl}
CFL = \frac{u\Delta t}{\Delta x}
\end{equation}

where $u$ is velocity at the point of interest, $\Delta t$ is time interval, and $\Delta x$ is grid spacing. 
$\Delta t$ and $\Delta x$ change with different temporal coarsening factors. 
We used a fixed $\Delta t$ for time integration throughout a simulation for both the reference solver and the learned solver. 
This was to emulate the advection solver in GEOS-Chem Classic which uses five minutes as the time interval for the finest nested grid simulation \cite{philip2016sensitivity}. 
In this work, we specified a cell mass and velocity at the cell center and flux output at the cell edge. 

This architecture was inspired by the design of finite-volume solvers, which also take as input the concentration and CFL information of grid cells in the neighborhood of the grid cell of interest (the ``stencil'') and output the flux between neighboring grid cells. 
Finite-volume methods enforce mass conservation by predicting mass flow between cells. 
As such, our learned CNN solver is designed to be mass conservative. 
However, in the 2-D applications in this paper described below, we clip negative concentration predictions to be zero, which can result in a slight increase in mass under certain conditions. 
This limitation (i.e., the violation of monotonicity) will be discussed in detail in the Results section. 

We constructed the convolutional neural network using three hidden layers, with each layer having 10 convolutional filters applied in parallel. 
The kernel size varies depending on the coarsening factor and layer number (Table~\ref{table:neighbor}). 
The first two layers used GeLU \cite{hendrycks2016gaussian} activation, chosen for its previous performance with physical systems \cite{kim2021stiff}. 
The last layer used hypertangent activation for output symmetry. 
The model outputs a prediction of mass flux between the grid cell of interest and its neighbor on the left side. 
We used array padding to specify zero-gradient Dirichlet boundary conditions for our CNN solver. 

For traditional numerical solvers, time step size is limited by the flow of information: local spatial gradients can only be calculated if a fluid parcel does not cross more than one grid cell boundary in a time step (i.e., the CFL threshold). 
As our machine-learned solvers are not directly approximating local spatial gradients, they are not subject to this limitation, but with increasing step sizes the ``information horizon'' within which grid cells can influence each others' behavior during a single time step expands. For example, with a CFL number of one, a given grid cell is influenced by one grid cell in each direction; with a CFL number of four, it is influenced by four grid cells in each direction. 
In our machine-learned solvers, we represent this expanding information horizon by increasing the kernel size with increasing temporal coarsening factor. 
As a kernel of size $k$ propagates information across $(k-1)/2$ grid cells in each direction, a three-layer CNN has an information horizon of $(k_1-1)/2 + (k_2-1)/2 + (k_2-1)/2 = (k_1+k_2+k_3-3)/2$, where $k_n$ is the kernel size of layer $n \in \{1,2,3\}$. 
When choosing the convolutional kernel sizes used in this analysis, we weighed tradeoffs between this theoretical information horizon and practical considerations regarding model speed and training time, both of which degrade with increasing kernel size. 
The kernel sizes we use here are enumerated in Table~\ref{table:neighbor}; as our goal is to evaluate the general characteristics of this approach rather than achieving the best-possible performance on a given benchmark, we did not extensively optimize the kernel sizes or any other hyperparameter used in this study. 

We used the Flux.jl machine learning software framework \cite{innes:2018} to train the convolutional neural network. 
After training, we transferred the Flux.jl model parameters to SimpleChains.jl \cite{simplechains}, a machine learning framework optimized for small neural networks \cite{simplechains_blog} to allow fast simulations. 

We normalize concentrations and CFL numbers by their absolute maximum values during a given timestep across the 1-D domain before feeding them into the CNN. 
We then multiply the CNN output vector by the input scaling factors (maximum concentration and absolute maximum CFL number), which we use as a representation of the flux between grid cells. 
The solver uses the flux output to calculate the next time step concentration by adding the difference between flux of the left and right edges to the concentration array as depicted in the right side of Figure~\ref{fig:solver}. 

\subsection{Data preparation}\label{subsec:data_prep}
To generate training, evaluation, and testing data, we implemented an advection solver used in the GEOS-Chem Classic atmospheric chemical transport model \cite{bey2001global} and in the NASA Goddard Earth Observing System (GEOS \cite{rienecker2008geos}; prior to transitioning to the cubed sphere grid \cite{putman2007finite}): the Piecewise Parabolic Method (PPM) \cite{colella1984piecewise} on a rectilinear latitude-longitude grid with multidimensional splitting suggested in \cite{lin1996multidimensional}. 
(Machine-learned solvers for the cubed-sphere grid used by recent versions of GEOS are an area for future work.) 
The PPM is a finite volume method which reconstructs subgrid cell scalar distributions with quadratic equations and computes flux on cell boundaries; details are available in the original paper \cite{colella1984piecewise}. 
As with the learned solver, our implementation of PPM uses the Julia programming language \cite{bezanson2012julia}. 

We used GEOS-FP wind fields \cite{geosfp} over the GEOS-Chem Classic North America domain, with an extent of $130\degree$W--$60\degree$W with a $0.3125\degree$ interval, and $9.75\degree$N--$60\degree$N with a $0.25\degree$ interval. 
To employ the finite volume method, we used an Arakawa C-grid \cite{arakawa1977computational} where the scalar values (i.e., concentration) are evaluated at centers of the grid cell and the velocity fields are evaluated at cell edges. 
As the GEOS-FP velocity fields that we use here (downloaded from \citeA{geosfpdata}) are on an Arakawa A-grid, we used linear interpolation to estimate velocities at edges where necessary using linear interpolation as implemented in the Julia library EarthSciData.jl v0.4.4 \cite{earthscidata}. 
Although this interpolation could introduce error in the velocity estimates, we expect this error to affect the predictions of the reference and machine-learned advection solvers similarly and therefore to not affect the results presented herein. 

To produce training datasets, we implemented 10-day-long 1-D advection simulations over the 202 latitudinal lines and 225 longitudinal lines laid on the North America domain described above. 
The time period of the simulation was from January 1st 00:00 UTC to January 11th 00:00 UTC in 2018 with a 5-minute time step. 
In each 1-D domain, we set the center ${1/3}$ to ${100}$~ppb and the other ${2/3}$ to ${0}$~ppb as an initial condition. 
This is the same strategy we used in our previous work \cite{park2024learned} to represent both the sharp initial concentration gradient and the relatively well-mixed concentration profile in the later part of the simulation, with the goal of ensuring training dataset diversity, to allow resulting models to represent even extreme events such as abrupt concentration increases owing to wildfires or chemical spills. 

We saved concentration values at the cell centers and the flux quantities at the cell edges for each time step. 
The total number of data points we used for training is (202 $\times$ 225 + 225 $\times$ 202) $\times$ 2880 = 261,792,000 when there is no temporal coarse-graining. 
We downsampled the dataset to sparse time domains that have 4$\times$, 8$\times$, 12$\times$, 16$\times$, and 32$\times$ larger time steps (and inverse-proportionally fewer data points). 
To downsample the data with ${\textit{k}}$ coarsening factor, we sampled the concentration in ${(\textit{k} \times n - \textit{k} + 1)^{th}}$ steps from ${n = 1}$ to ${\textit{k} \times n - \textit{k} + 1 \leq 2880}$. 
We accumulated the flux value for a specific edge across ${k}$ time steps to ensure mass conservation. 
(i.e., the machine learning solver was trained to give the sum of flux over multiple time steps rather than giving a single flux quantity in a specific time step.) 
We verified that the simulation of concentrations with this flux accumulation method reproduced the time evolution of concentrations present in the original dataset to within machine precision. 

To produce the testing datasets we implemented 2-D advection over same spatial and temporal domain as a reference. 
To do this, we performed the non-directional splitting of the latitudinal and longitudinal advection as described in detail in \citeA{lin1996multidimensional}, the essence of which is in Equation~\ref{eq:splitting}:
\begin{equation}\label{eq:splitting}
Q^{n+1} = Q^{n} + F(Q^{n}+\frac{1}{2}G(Q^{n})) + G(Q^{n}+\frac{1}{2}F(Q^{n}))
\end{equation}
where $Q^n$ is the scalar value in a given cell center at $n^{th}$ time step, and $F$ and $G$ are the (identical) latitudinal and longitudinal direction advection operators, respectively. 
For the initial condition of the 2-D simulation, we set the center ${1/9}$ of the spatial domain (i.e., the center ${1/3}$ of the domain in both North-South and East-West directions) to ${100}$~ppb while the remaining ${8/9}$ was zero. 

\subsection{Model training and evaluation}\label{subsec:method_train_eval}

In each training iteration, we input five time steps of 1-D advection sampled at a randomly chosen starting time and a randomly sampled latitude or longitude. 
We introduced the concentration and CFL value to the CNN-based flux solver and the solver output the flux terms using the current neural network parameters. 
By integrating the flux output the solver calculated the next time step concentration. 
We used the calculated concentration to predict the flux in the next time step recurrently for five time steps. 
We evaluated the mean absolute error between predicted fluxes during the five-time-step integration and the original fluxes from the training dataset. 
We trained the model over multiple time step rollouts as proposed by \citeA{zhuang2021learned} to promote consistency and the stability of long-term simulations. 
We alternated training samples drawn from the latitudinal and longitudinal directions and shuffled the order of the training data for each epoch. 
We used a learning rate that decays as training progresses; the learning rate ($\eta$) in each training epoch is shown in Equation~\ref{eq:decay}:
\begin{equation}\label{eq:decay}
\eta = \frac{2 \times 10^{-3}}{1+\rm epoch}.
\end{equation}

After each epoch we evaluated our solver against 10-day-long 2-D advection results. 
The wind datasets we used to generate 1-D training dataset and 2-D testing dataset are identical. 
Still, the testing dataset represents unseen conditions to the solver because: 1) each training iteration feeds only one latitudinal and longitudinal line from 202 and 225 lines each with only five time steps, 2) the concentration input streams are not the same as 1-D training dataset, and 3) 2-D splitting inputs half the time step used in training multiple times, which effectively halves the CFL number input into the CNN. 
We fed the same initial condition and velocity field (i.e., CFL) used to reproduce the 2-D testing dataset using the trained solver; for 2-D splitting we used Equation~\ref{eq:splitting}. 
We used mean absolute error, root mean square error, and $r^2$ to quantify model performance. 
To choose the final models we present below in the Results section, we selected model parameters saved at the training iteration gave the best overall performance on these metrics on this 2-D testing dataset. 
When the best set of parameters based on MAE, RMSE, and $r^2$ did not match we selected the model parameters with lowest discrepancy among those statistics. 
Our intention in using 2-D testing rather than the 1-D testing was to evaluate the usefulness of our 1-D solver in a multidimensional application, which is the ultimate purpose of the solver. 

\subsection{Generalization tests}
As described above, we trained our model using January 2018 ground level wind data in North America in January. 
To evaluate the robustness of the model against out-of-distribution input data, we performed two generalization tests using velocity fields from different seasons and different vertical levels. 
We used these tests to evaluate solver robustness in handling unseen wind characteristics and in representing the temporal evolution of concentration profiles that are substantially different from the training dataset. 

For the seasonal generalization test, we ran 10-day-long 2-D advection using the wind data starting the first day of every month in 2018. 
For the vertical level generalization test, we evaluated the 2-D integration using the wind field at each of the 72 vertical levels spanning the vertical domain of the GEOS-FP wind data product, up to 78.146 km (0.015 Pa) in altitude \cite{geosfpverticalgrid}, using January 1st 2018 for the starting date of the simulation. 
The generalization testing protocol was the same as the 2-D testing described in Section~\ref{subsec:method_train_eval}.

For both generalization tests, the learned solver needed to address larger CFL values than the maximum CFL number seen in training. 
In these cases, we divided the CFL numbers by the maximum CFL number in training dataset, and repeat the 1-D advection with smaller time interval to guarantee the input CFL number fell within the range represented in the training data. 
We performed the time slicing only for the generalization testing. 

\subsection{Timing}\label{subsec:timing}
We evaluated the computational time required to run 10-day-long 2D advection simulations using the reference solver and the learned solvers, both using various time coarsening factors. 
We only measured simulation time, excluding overhead such as loading time for required packages, velocity data, and array allocation. 
To solely evaluate the computing time, we ensured that neither the reference solver nor the machine-learned solver performed any memory allocations. 
We used a single CPU core (Intel 6248 Cascade Lake) within an HPE Apollo 6500 node installed in the University of Illinois Campus Cluster. 
We used BenchmarkTools.jl \cite{benchmarktools} to procure a robust evaluation of computational time over multiple trials. 
We calculated the relative speed of the learned solver by dividing the computational time taken by the learned solver by the time taken by the reference solver in the original time resolution. 
To enable comparisons between our temporally-coarsened learned solver and a temporally coarsened reference solver, we also evaluated accuracy and computational time for the reference solver at different levels of temporal coarsening.

\section{Results}\label{sec:results}

\subsection{Solver performance in 2-D advection}
\begin{figure}
\includegraphics[width=1.0\textwidth]{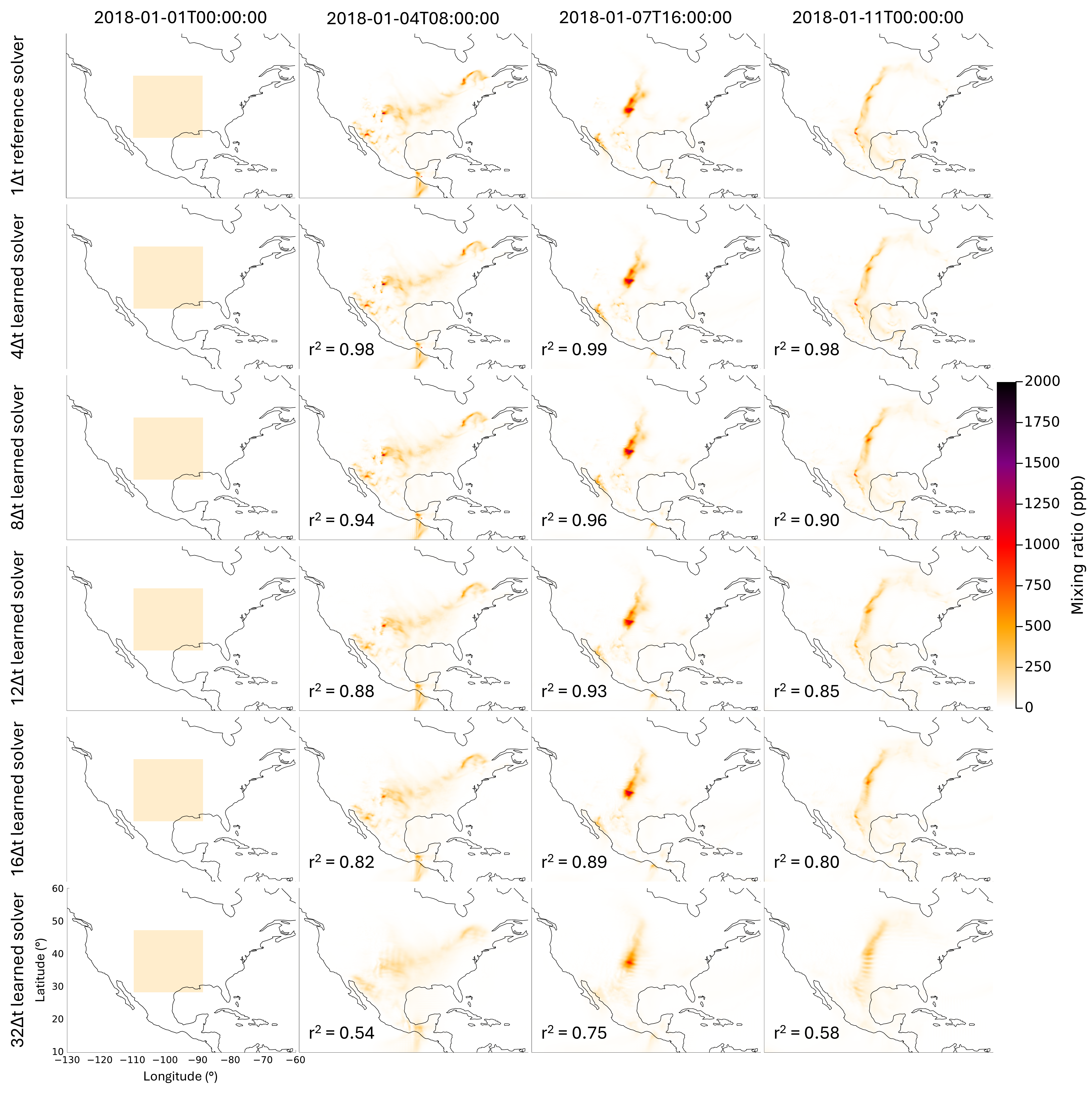}
\caption{Time series representation of numerical advection using the reference solver (top row) and the learned solvers in different time coarsening factors (second to sixth rows).}
\label{fig:time_series}
\end{figure}

Figure~\ref{fig:time_series} shows the temporal evolution of the concentration profiles over the 10-day-long ground level advection simulation. 
The top row of Figure~\ref{fig:time_series} shows the simulation results using the reference advection scheme; the other rows show the results for the machine learned flux solver with temporal coarse-graining.
(All panels in Figure~\ref{fig:time_series} contain maximum concentrations that do not monotonically decrease with time. This is a non-physical artifact caused by our simulation of only two dimensions of a three-dimensional flow field, but affects both reference and ML solvers in the same manner and so is not an important aspect of our evaluation.) 

All learned coarse solvers maintain the major plume shape in every temporal resolution tested, however solvers with larger temporal coarsening result in relatively blurry concentration distributions, which would be indicative of numerical dissipation in a traditional solver. 
The overall concentration ranges simulated by solvers with larger coarsening factors are lower than the baseline simulation, which would also be indicative of numerical dissipation. 

In the former case, error could be improved for example through including a spectral term in the training loss function, to encourage the power spectrum of the learned solver simulations to match those of the reference simulations.
We will leave this for future work. 
In the latter case, it may be possible to improve accuracy through hyperparameter tuning, such as increasing the filter size or number of layers, but exhaustively maximizing accuracy on a particular benchmark is not the goal of this analysis. 
Instead, we aim here to establish a baseline of performance across multiple benchmarks without extensive tuning in order to avoid the effects of Goodhart's Law \cite{manheim2018categorizing, goodhart1984problems}.

\begin{figure}
\includegraphics[width=1.0\textwidth]{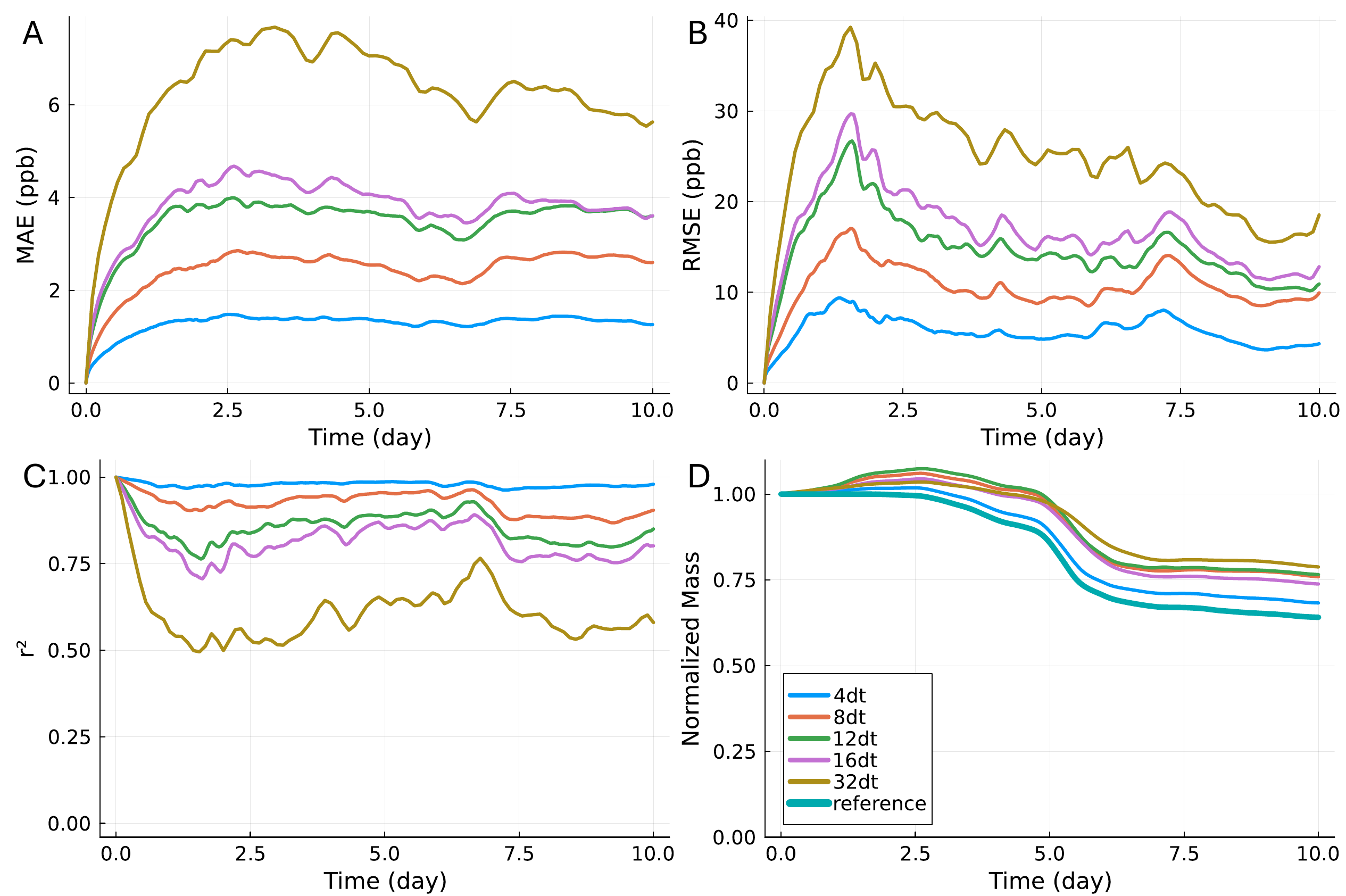}
\caption{The evolution of spatial statistics over time. A: mean square error, B: root mean square error, C: r$^2$, and D: normalized sum of the concentration over the spatial domain. Although our learned solver conserves mass by design, we clip negative concentration predictions to zero, resulting in the slight overall mass increases shown at the beginning of the simulation in Panel D. 
Mass exiting the spatial domain through the lateral boundaries causes the loss in mass occurring later in the simulation.}
\label{fig:error_stats}
\end{figure}

Figure~\ref{fig:error_stats} visualizes the evolution of error statistics during the simulation time period in each of the temporal coarse-graining factors studied. 
Larger coarsening factors result in larger errors and lower $r^2$ values. 
For all temporal coarsening factors, the error increases in the early phase until around 2.5 days and then stabilizes or slightly decreases, possibly because the sharp gradient in the initial concentrations is more difficult to simulate than are the more gradual gradients found later in the simulation. 
$r^2$ values show similar patterns. 

The normalized total mass over the spatial domain is shown in Figure~\ref{fig:error_stats}D. 
The learned solver at every time coarsening factor shows a slight mass increase at the beginning of the simulation, which turns to a decrease after 2.5 days. 
(The reference solver also experiences a decrease in mass; in both cases mass decreases are caused by mass leaving the domain through the lateral boundaries.) 
Mass increase occurs when the neural network predicts concentration changes that would result in negative concentrations in the next time step, and the negative concentrations are clipped to zero by the positivity limiter, but the positivity limiter does not cancel out corresponding concentration increase in the neighboring grid cell. 
Therefore, every time a negative concentration prediction is clipped to zero, there is a net increase in the overall mass in the system. 
Methods to mitigate this violation of monotonicity are an area for future work.

Power spectrum analysis shows that the frequency distribution of the learned solver is similar to that of the reference solver (Figure~\ref{fig:spectral}). 
The 4$\times$ and 8$\times$ temporally-coarsened learned solver shows nearly identical spectral power densities with the reference solver. 
The other learned solvers result in power densities with similar shape to the reference solver but with but lower amplitudes. 
The lower amplitudes with larger coarsening factors are consistent with the numerical dissipation effects posited above, and could potentially also be fixed using the approach suggested above. 

\subsection{Generalization capability}\label{subsec:general}
\begin{figure}
\includegraphics[width=0.75\textwidth]{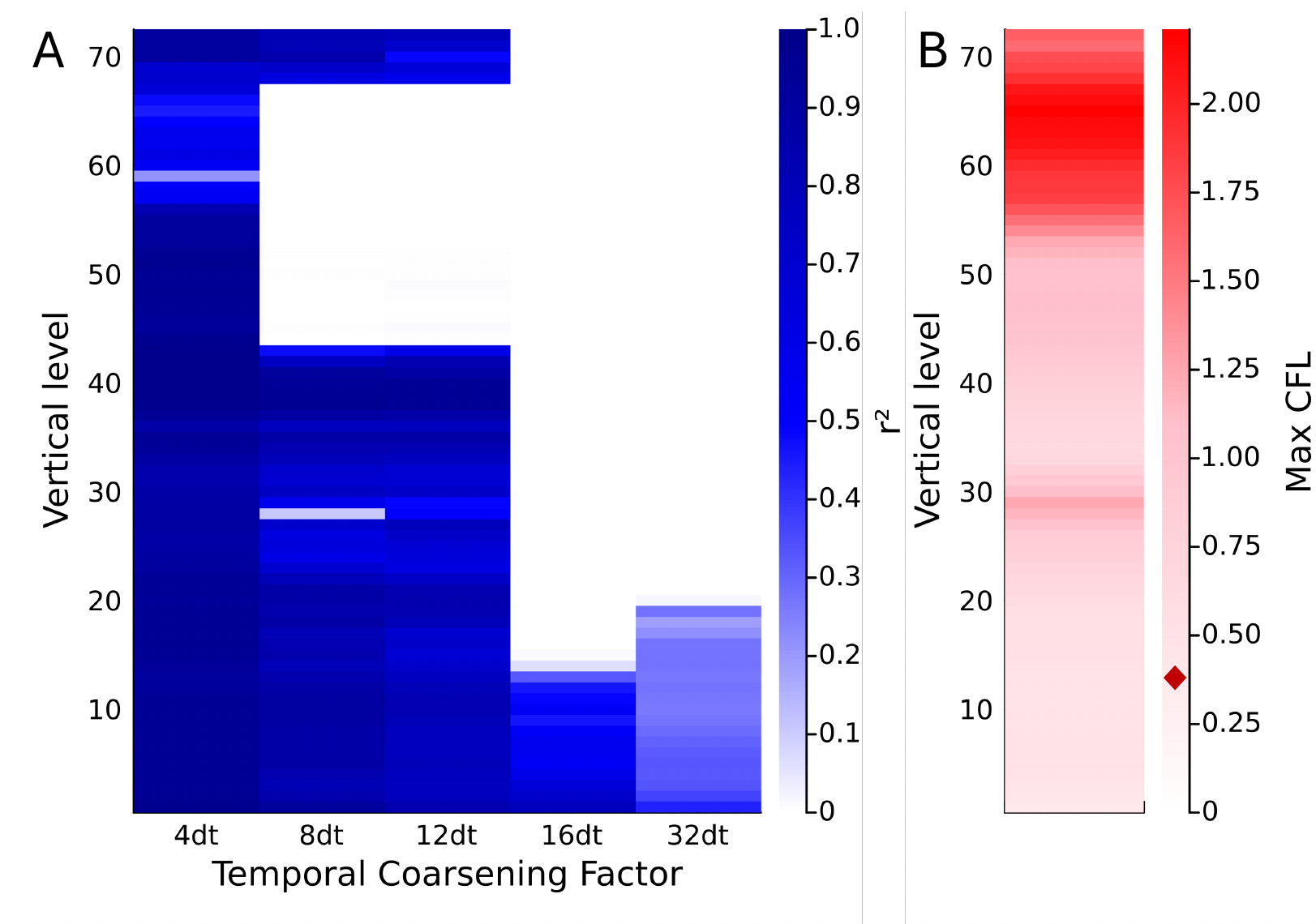}
\caption{Results of generalization tests of our machine-learned advection solver against every GEOS-FP vertical level for January 2018. 
A) r$^2$ values for 10-day-long 2-D advection simulations. 
B) Maximum CFL values for each simulation. 
The red diamond in the colorbar indicates the maximum CFL value in the training dataset. 
(These CFL values can be multiplied by the temporal coarsening factors to obtain the effective CFL values for the learned solvers.)}
\label{fig:gen_test_level}
\end{figure}

Figure~\ref{fig:gen_test_level}A shows the r$^2$ values of the solvers for the 72 vertical layers in the GEOS-FP vertical grid, spanning up to $\sim$80~km in elevation. 
The 4$\times$-temporal-coarsened solver showed good generalization performance in every vertical level tested. 
The 8$\times$-, 12$\times$-, and 16$\times$-coarsened solvers showed good generalization capability until at around 40$^{th}$ level, which is approximately 21~km above sea level. 
After the 40$^{th}$ level, the models with larger coarsening factors tended to fail to produce numerically stable simulations. 
The 32 times coarse solver showed limited generalization up until 10th vertical level, which is $\sim$1.3~km above ground level. 

\begin{figure}
\includegraphics[width=0.75\textwidth]{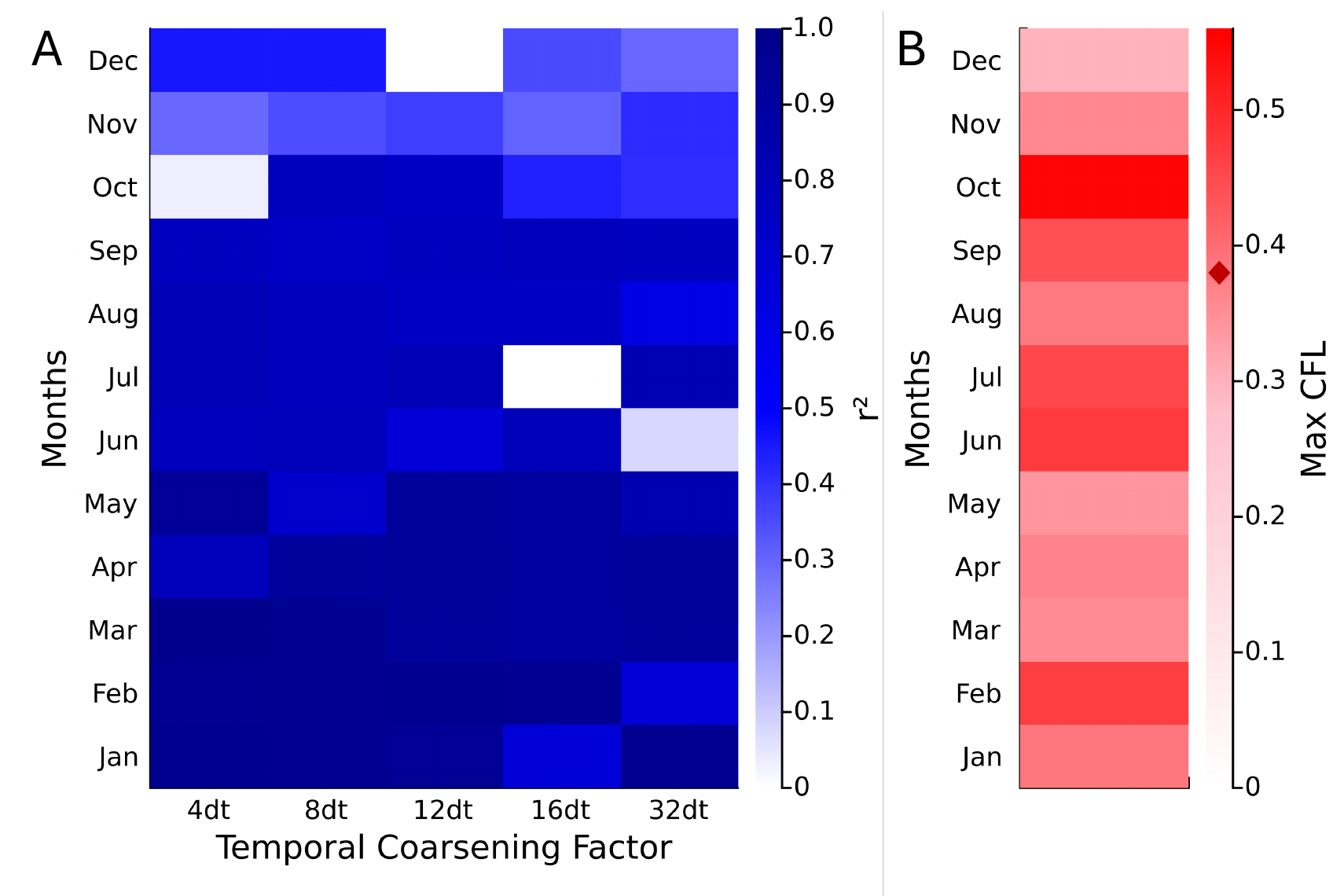}
\caption{Results of generalization tests against ground level simulations for each month of 2018. A) r$^2$ values for 10-day-long 2-D advection simulations. B) Maximum CFL values for each simulation. The red diamond in the colorbar indicate the maximum CFL value in the training dataset. (These CFL values can be multiplied by the temporal coarsening factors to obtain the effective CFL values for the learned solvers.)}
\label{fig:gen_test_month}
\end{figure}

Figure~\ref{fig:gen_test_month}A shows the r$^2$ values learned solver performance in 10-day-long advection simulations using ground level wind data from every month of 2018. 
Solver performance degrades in June and October, resulting in unstable simulations. 
For the remaining months, the solvers maintain similar performance to that observed for the training month of January. 
This suggests the good generalization capability of the solvers in resolving seasonality, although there are issues with numerical stability in specific cases. 
We expect that including different seasons and vertical levels in our training data would substantially improve performance on these tests; we do not do so here because our goal is to test the robustness of our method rather than optimizing it for production use.

We investigated whether the limits to the generalization capability described above are associated with wind speed. 
Figure~\ref{fig:gen_test_month}B shows the maximum CFL value in each month in our generalization testing. 
The maximum CFL value in the training dataset (i.e., ground level wind on January 2018) is 0.386 and marked with a red diamond in Figure~\ref{fig:gen_test_month}B. 
The largest monthly maximum CFL value occurred October (0.552), followed by June (0.475). 
October was the month when the learned solvers in 8, 12, and 16 times temporal coarsening factor generated numerically unstable simulations. 
For June, the learned solver with a 12 times temporal coarsening factor failed to produce a numerically stable simulation. 
Figure~\ref{fig:gen_test_level}B shows the maximum CFL value in every 72 vertical layer in the testing time period, showing that the vertical levels where the learn solver failed also usually (but not always) had relatively high CFL values. 
Together, this information suggests the learned solvers often fail in generating numerically stable simulations for CFL values substantially larger than the CFL values represented in the training dataset, and therefore it may be important for learned solvers intended for operational use to be trained on data including a wide range of CFL values. 
However, as some vertical levels with high maximum CFL values were able to be successfully simulated with our solvers, future work identifying additional metrics important for solver performance could be useful.


\subsection{Speedup gain by the learned coarse solver}\label{subsec:speedup}

\begin{figure}
\includegraphics[width=0.5\textwidth]{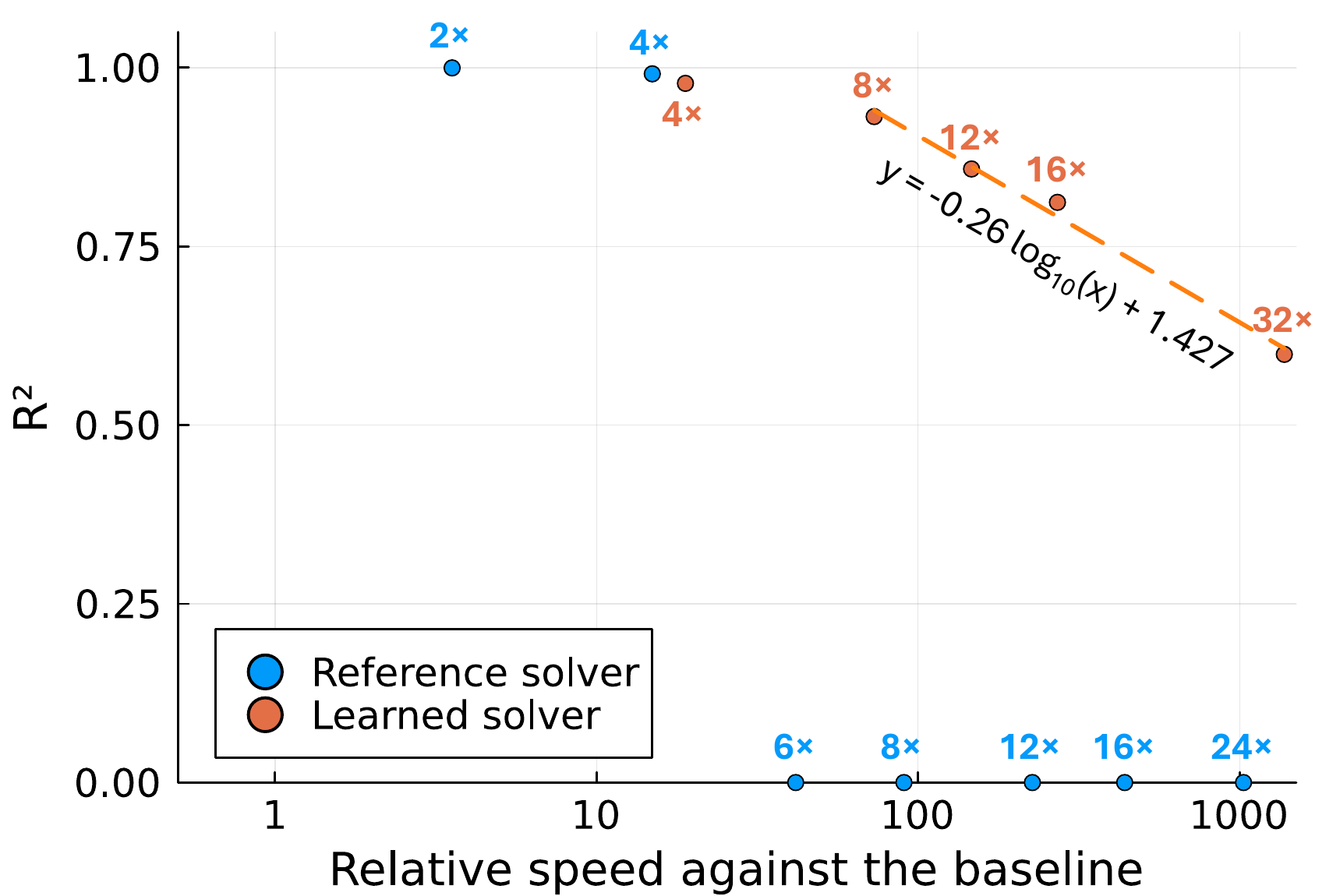}
\caption{Precision versus speed for the reference PPM solver and the machine learned flux solver with different temporal coarsening factors. The numbers annotating the points indicate temporal coarsening factors.}
\label{fig:speedup}
\end{figure}

Figure~\ref{fig:speedup} shows the speed and accuracy of our learned solvers and the reference solver at different levels of temporal coarsening. 
The reference solver is able to operate at up to 4$\times$ temporal coarsening, achieving a $12\times$ speedup without substantial loss in accuracy.
However, at larger coarsening factors, the reference model continues to run faster but is not numerically stable (because CFL $>$ 1) and therefore does not produce useful results. 
The machine-learned solvers achieve a similar level of accuracy and speed gain at 4$\times$ temporal coarsening as the reference solver does, but in contrast to the reference solver they yield gradually degrading accuracy as the coarsening factor increases, achieving in the extreme case a 92.43$\times$ speedup compared to the fastest viable reference model simulation while maintaining an $r^2$ of 0.599.

The correlated relationship between speed and accuracy of the learned solver sets up a paradigm where users could make tradeoffs between solver speed and accuracy depending on the requirements of their use case. 
Simple linear regression between the log-transformed relative speed and R$^2$ reveals users of the solvers we have developed here could gain a 10-fold speedup for every 0.26 decrease in R$^2$ they are willing to give up (in the range from 8$\times$ to 32$\times$ temporal coarsening). 
Users could choose a lower factor of temporal coarsening (or the reference solver) when high accuracy is desired. 
There may also exist tradeoffs between spatial and temporal resolution such that the use of temporal coarsening could allow spatial refinement, increasing spatial precision and possibly overall accuracy for some metrics \cite{paolella2018effect}. 
This could be an area for exploration in future work.

The computational performance shown in Figure~\ref{fig:speedup} was made possible through the use of a machine learning framework specialized for small neural networks—SimpleChains.jl \cite{simplechains}.
When using a more traditional neural network software library (Flux.jl \cite{innes:2018}), we are only able achieve a maximum of $3.64\times$ speedup against the fastest viable reference solver (Figure~\ref{fig:speedup_SI}) as compared to the $92.43\times$ maximum speedup shown in Figure~\ref{fig:speedup}.

We do not consider computational speedups that could be achieved through the use of hardware accelerators such as GPUs because both the reference and learned solvers could be accelerated through the use of a GPU. 

\section{Discussion}\label{sec:discussion}

The major achievement of this study is the demonstration of a 19$\times$ to 1377$\times$ speedup in the simulation of numerical advection while maintaining the original spatial resolution. 
This resolves a major limitation of previous work \cite{park2024learned} where the learned solvers in the original spatial grid were not able to produce stable simulations. 
The ability to increase computational speed without decreasing spatial resolution is relevant to applications of advection modeling, as a core research and development goal is often to increase spatial resolution \cite{national2020earth, ncep2023proj} rather than decrease it. 
Given this importance of spatial resolution in air quality modeling, this work represents a milestone in the path toward an operationally usable machine-learned advection solver that can accelerate air quality model simulations.

In the work described above, we have focused on describing a novel machine-learned advection solver architecture which incorporates physical principles with the goal of maximizing computational acceleration and the range of conditions under which the solver can be used, while minimizing the amount of training data required to parameterize it. 
We believe that the solver design described above represents the state of the science in its ability to meet these design goals. 
However, there remain areas for improvement, including performance under CFL conditions (in the non-temporally-coarsened data) which exceed those represented in the training data, as well as our model's propensity to predict negative concentrations and the monotonicity violation that results when those negative predictions are clipped to zero. 
Additionally, as the current study has not focused on producing an operationally-usable advection operator, doing so is a clear area for future work. 
Creating a solver with the goal of operational use would require training the model on a larger amount and diversity of data and searching for a combination of hyperparameters—for example kernel size; number of model layers, filters, and kernels per layer; loss function choice; input normalization methods; and learning rate schedule—which provide optimal tradeoffs between and speed and accuracy, and would also require extensively testing solver performance in operational settings. 
An additional potential area for future work could be the representation of vertical advection, but as CFL values are often much lower for vertical advection than they are for horizontal advection models, this may not be a priority area for future research.

We have demonstrated the application of our solvers in the context of passive mass transport, which is a core component of atmospheric chemical transport modeling. 
However, as the transport of heat and momentum are analogous to the transport of mass, the results here would be readily transferable to those applications. 
Additionally, if trained on data generated using reference models with detailed turbulence representations (such as large eddy simulations or direct numerical simulations), our learned models could provide implicit turbulence closures by reproducing the emergent effects of turbulence without simulating it in detail. 
These are all promising areas of future research. 
There already exists an extensive literature proposing ML representations of the full Navier Stokes equations (rather than just advection as we study here) \cite{vinuesa2022enhancing, kochkov2021machine, mcgreivy2024weak, taira2025machine}, but physics-informed machine learning for atmospheric fluid mechanics in particular has been less-well studied \cite{mojgani2023extreme, cheng2022deep}, and approaches derived from the work herein could be applied in that field.

\section*{Open Research Section}
All the code in this study to produce datasets and evaluate the solver function, and the neural network parameters are uploaded in a GitHub repository \cite{mladv}. 

\acknowledgments
This work is supported by an Early Career Faculty grant from the National Aeronautics and Space Administration (grant no. 80NSSC21K1813), Assistance Agreement RD-84001201-0 awarded by the U.S. Environmental Protection Agency and Grant NSF RISE 24‐25760 from the U.S. National Science Foundation. 
It has not been formally reviewed by EPA, NASA or NSF. 
This work made use of the Illinois Campus Cluster, a computing resource that is operated by the Illinois Campus Cluster Program in conjunction with the National Center for Supercomputing Applications (NCSA) and which is supported by funds from the University of Illinois at Urbana-Champaign. 
The views expressed in this document are solely those of authors and do not necessarily reflect those of funding agencies. 
EPA does not endorse any products or commercial services mentioned in this publication. 
MP received support from the Carver Fellowship and Illinois Distinguished Fellowship.

%
%

\bibliography{references}

%
%
%
%
%

\clearpage
\section*{Supporting Information}

\setcounter{table}{0}
\renewcommand{\thetable}{S\arabic{table}}

\begin{table}[h!]
\centering
\caption{Temporal coarsening factor, kernel size in each hidden layer of the model, and resulting information horizon for each time step.}
\begin{tabular}{c c c c c}
\hline
\makecell{\textbf{Coarsening}\\\textbf{Factor} \\} & 
\makecell{\textbf{First Layer}\\\textbf{Kernel Size}\\ ($k_1$)} & 
\makecell{\textbf{Second Layer}\\\textbf{Kernel Size}\\ ($k_2$)} & 
\makecell{\textbf{Third Layer}\\\textbf{Kernel Size}\\ ($k_3$)} &
\makecell{\textbf{Information Horizon}\\\textbf{(in each direction)}\\ $(k_1 + k_2 + k_3 -3)/2$} \\
\hline
4 & 3 & 3 & 3 & 3\\
8 & 5 & 3 & 3 & 4\\
12 & 5 & 5 & 5 & 6\\
16 & 5 & 5 & 5 & 6\\
32 & 7 & 7 & 5 & 8\\
\hline
\end{tabular}
\label{table:neighbor}
\end{table}

\clearpage

\setcounter{figure}{0}
\renewcommand{\thefigure}{S\arabic{figure}}

\begin{figure}
\includegraphics[width=1.0\textwidth]{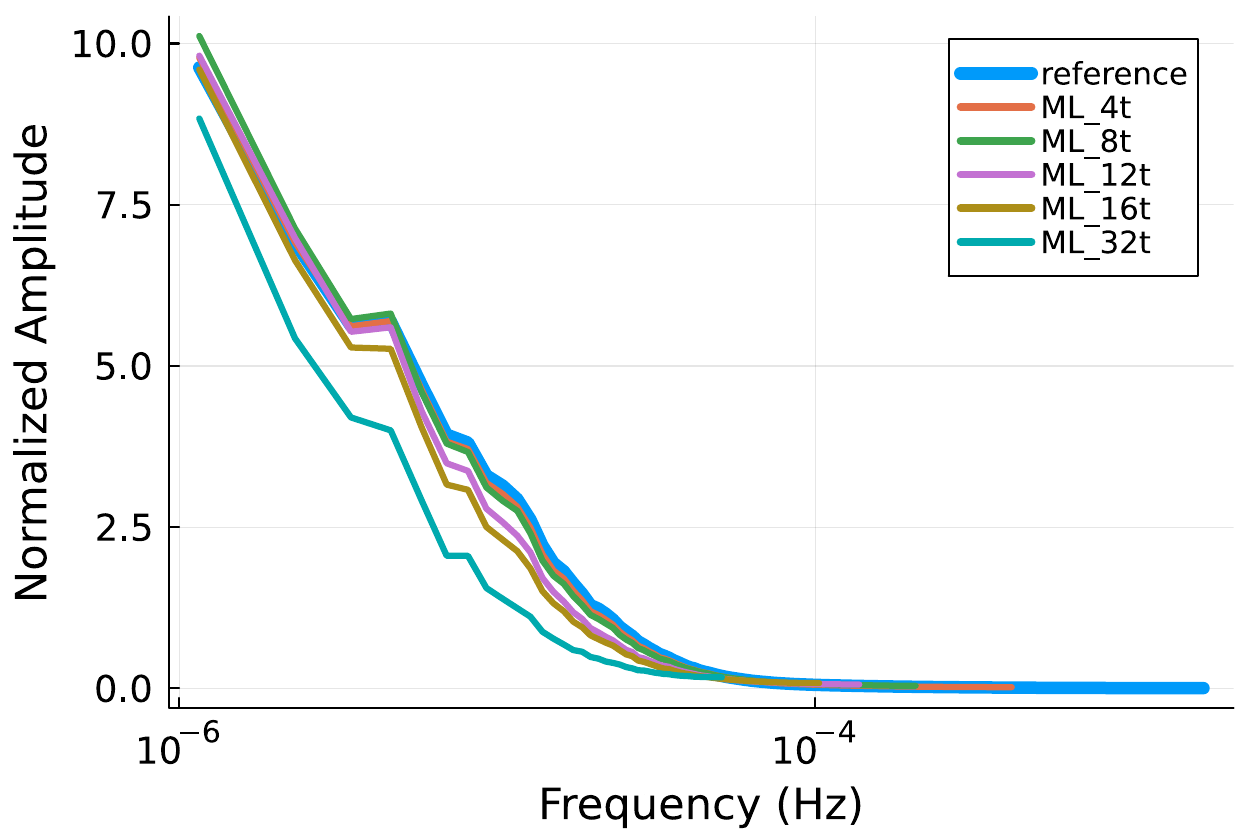}
\caption{Spectral power analysis of the advection simulation results with the reference solver and the learned temporal-coarse-graining solvers.}
\label{fig:spectral}
\end{figure}

\begin{figure}
\includegraphics[width=1.0\textwidth]{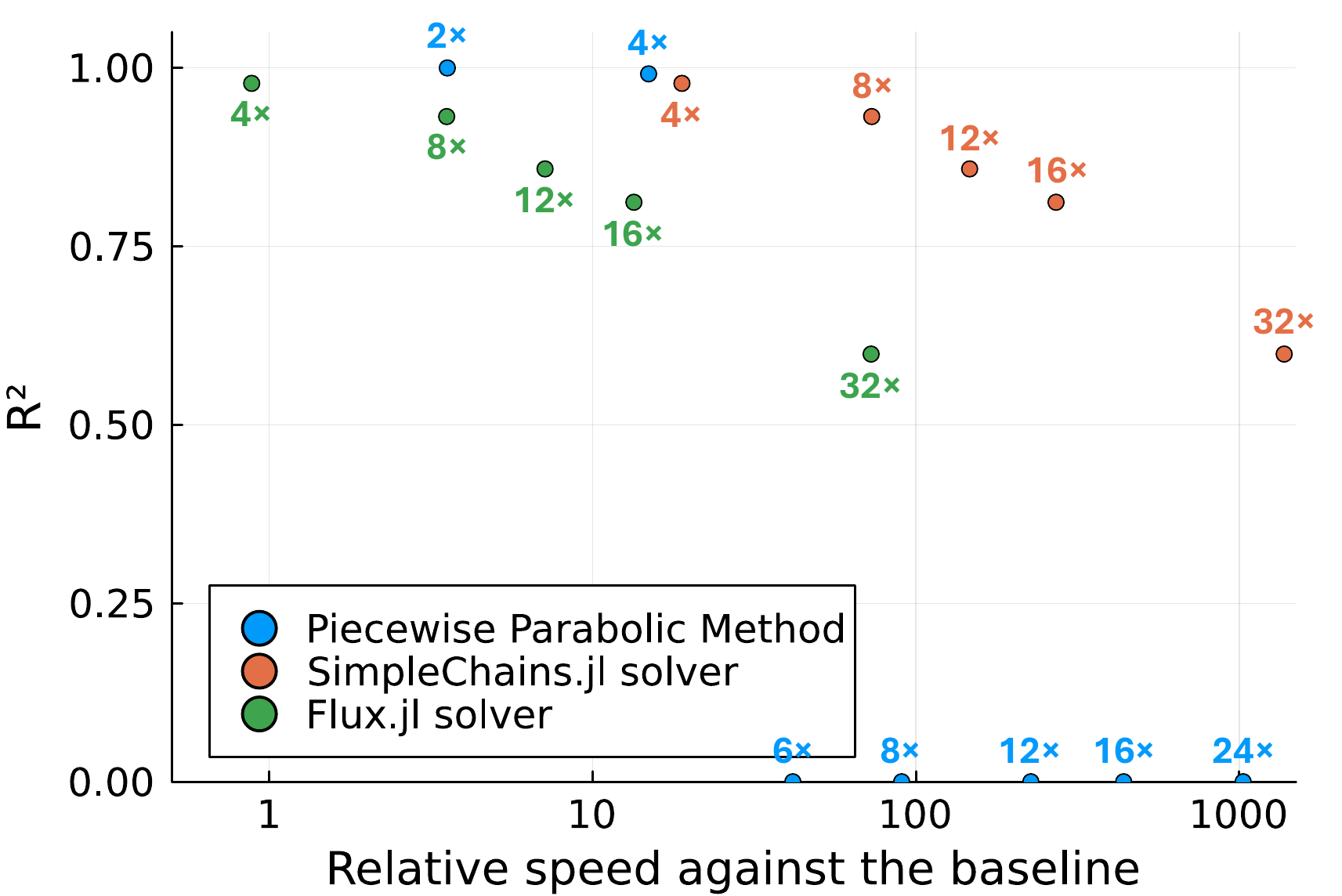}
\caption{Precision versus speed for the reference PPM solver and the machine learned flux solvers with different temporal coarsening factors. The SimpleChains.jl solver and Flux.jl solver shares the same parameter set while SimpleChains.jl gained higher speedup. The SimpleChains.jl and Flux.jl neural networks use the same parameters so predictive performance is the same in the machine precision level between the two solvers—only computational performance differs. The numbers annotating the points indicate the temporal coarsening factors.}
\label{fig:speedup_SI}
\end{figure}


\end{document}